\documentclass[aps, prd, superscriptaddress, preprintnumbers]{revtex4}

\usepackage{graphicx,amsfonts,amsmath,amssymb,amstext}
\usepackage{float,wrapfig}
\usepackage{subfigure, psfrag}
\usepackage{dsfont}
\usepackage{color}
\usepackage{bm}

\newcommand{\be}{\begin{equation}}
\newcommand{\ee}{\end{equation}}
\newcommand{\ba}{\begin{eqnarray}}
\newcommand{\ea}{\end{eqnarray}}

\definecolor{purple}{rgb}{0.8,0,0.6}
\definecolor{darkgreen}{rgb}{0.00,0.6,0.00}
\begin{document}

\title{Hydrodynamics of Fermi arcs: Bulk flow and surface collective modes}
\date{April 20, 2019}

\author{E.~V.~Gorbar}
%\email{gorbar@bitp.kiev.ua}
\affiliation{Department of Physics, Taras Shevchenko National Kiev University, Kiev, 03680, Ukraine}
\affiliation{Bogolyubov Institute for Theoretical Physics, Kiev, 03680, Ukraine}

\author{V.~A.~Miransky}
%\email{vmiransk@uwo.ca}
\affiliation{Department of Applied Mathematics, Western University, London, Ontario, Canada N6A 5B7}

\author{I.~A.~Shovkovy}
%\email{igor.shovkovy@asu.edu}
\affiliation{College of Integrative Sciences and Arts, Arizona State University, Mesa, Arizona 85212, USA}
\affiliation{Department of Physics, Arizona State University, Tempe, Arizona 85287, USA}

\author{P.~O.~Sukhachov}
%\email{psukhach@uwo.ca}
\affiliation{Department of Applied Mathematics, Western University, London, Ontario, Canada N6A 5B7}

\begin{abstract}
The hydrodynamic description of the Fermi arc surface states is proposed. In view of the strong suppression
of scattering on impurities, the hydrodynamic regime for Fermi arc states should be, in principle, plausible.
By using the kinetic theory, the Fermi arc hydrodynamics is derived and the corresponding effects on the
bulk flow and surface collective modes are studied. For the bulk flow, the key effect of the proposed Fermi
arc hydrodynamics is the modification of the corresponding boundary conditions. In a slab geometry, it is
shown that, depending on the transfer rates between the surface and bulk, the hydrodynamic flow of the
electron fluid inside the slab could be significantly altered and even enhanced near the surfaces.
As to the spectrum of the surface collective modes, in agreement with earlier studies, it is found that
the Fermi arcs allow for an additional gapless spectrum branch and a strong anisotropy of the surface
plasmon dispersion relations in momentum space. The gapped modes are characterized by closed elliptic contours of constant frequency in momentum space.
\end{abstract}

\maketitle

\section{Introduction}
\label{sec:Introduction}

Weyl semimetals are materials with a relativisticlike energy spectrum in the vicinity of isolated
Weyl nodes in the Brillouin zone. (For recent reviews on Weyl semimetals, see
Refs.~\cite{Yan-Felser:2017-Rev,Hasan-Huang:rev-2017,Armitage-Vishwanath:2017-Rev}.)
The nodes have nonzero topological charges with the monopolelike Berry curvature \cite{Berry:1984}
and always occur in pairs of opposite chirality \cite{Nielsen-Ninomiya-1,Nielsen-Ninomiya-2}.
In each pair, the Weyl nodes can be separated in energy and/or momentum, which indicates
breaking of the parity-inversion (PI) and/or time-reversal (TR) symmetries, respectively. The nontrivial
topology and the relativisticlike nature of quasiparticles also affect the transport properties of Weyl
semimetals, e.g., leading to a negative longitudinal magnetoresistivity that was first predicted in
Ref.~\cite{Nielsen}. (For recent reviews of the transport phenomena, see
Refs.~\cite{Lu-Shen-rev:2017,Wang-Lin-rev:2017,Gorbar:2017lnp}.)

The nontrivial bulk topology of Weyl semimetals is also reflected in unusual surface states known
as the Fermi arcs \cite{Savrasov:2011}. Unlike surface states in ordinary materials, the Fermi arcs
form open segments in momentum space that connect Weyl nodes of opposite chirality
\cite{Savrasov:2011,Haldane:2014}.
The surface states in Weyl semimetals were first observed via the angle-resolved photoemision spectroscopy~\cite{Bian,Qian,Yang-Chen:2015,Xu-Hasan-NbAs:2015,Xu-Hasan-TaP:2015,Xu-Shi:2016,Hasan-Huang:rev-2017} and reconfirmed later by the observation of the quasiparticle interference patterns \cite{QPI:Hasan-2,QPI:Batabyal,Inoue-Yazdani-QIP:2016,Gyenis-Yazdani-QIP:2016}. It is important to note that the energy
dispersion of the Fermi arc states is effectively one dimensional and linear (see, e.g., Ref.~\cite{Murakami:2014}).
This may suggest that their transport properties are similar to that of the one-dimensional chiral
fermions and should be nondissipative. However, as we showed in Ref.~\cite{Gorbar:2016aov},
the Fermi arc transport is, in fact, dissipative because of the scattering between the surface and
bulk states in Weyl semimetals. The dissolution of Fermi arcs in the presence of strong disorder
was also confirmed numerically in Refs.~\cite{Slager:2017tgx,Wilson-DasSarma:2018}.

Electronic collective excitations provide additional powerful probes of the nontrivial properties of
Weyl semimetals. The topology is imprinted, for example, in anomalous helicons \cite{Pellegrino:2015},
surface plasmon polaritons \cite{Hofmann:2016,Tamaya:2018}, chiral magnetic plasmons
\cite{Gorbar:2016ygi,Gorbar:2016sey,Long:2017xnj}, etc. The effect of the Fermi arcs on the
surface plasmons was studied in Refs.~\cite{Song:2017,Andolina:2018,Losic:2018}. The authors
of Ref.~\cite{Song:2017} employed a simple phenomenological model valid in the long-wavelength
limit. The hybridization of the Fermi arc states and conventional surface plasmons is controlled via the
anomalous Hall conductivity and a phenomenologically included Drude weight. Further, the surface
plasmon excitation spectrum in Weyl semimetal within the random-phase approximation was
determined in Ref.~\cite{Losic:2018}. The treatment of the surface plasmons in Ref.~\cite{Andolina:2018}
is more sophisticated and is based on the direct quantum-mechanical calculations. Despite the
difference in their approaches, studies in Refs.~\cite{Song:2017,Andolina:2018}
predict the open hyperbolic constant-frequency contours for the surface plasmons. The nontrivial
patterns of the surface plasmons can be measured by the scattering-type near-field optical
spectroscopy (for a recent review, see Refs.~\cite{NovotnyStranick:2006,Basov-rev:2016}) as well as the
momentum-resolved electron energy-loss spectroscopy (see e.g., Refs.~\cite{Wang:1995,Abajo:2010} and the references
therein). Experimentally, the electron energy loss in Weyl semimetals was recently studied in
Ref.~\cite{Chiarello:2018}.

Since Weyl semimetals are typically characterized by low impurity scattering rates (see, e.g.,
Refs.~\cite{Armitage-Vishwanath:2017-Rev,Wang-Lin-rev:2017,Zhang-Hasan-TaAs:2016,Wang-Xu-NbP:2016,LiChenJin2016}
for the scattering rates and crystal quality estimations), one might suggest that a hydrodynamic regime
of electron transport could be eventually realized in many such materials. Originally, the
possibility of such a regime for charge carriers in sufficiently clean solids was discussed in the pioneering
papers by Gurzhi~\cite{Gurzhi:1963,Gurzhi:1968}. Electron hydrodynamics requires that the electron-electron scattering rate
dominates over the electron-impurity and electron-phonon ones. Recently, such a regime was
experimentally confirmed in the Weyl semimetal tungsten diphosphide (WP$_2$)~\cite{Gooth:2017},
where the characteristic quadratic dependence of the electrical resistivity on the cross section of the
wire as well as a strong violation of the Wiedemann--Franz law were observed.

Theoretically, the nontrivial topological properties of Weyl semimetals, connected with the energy and
momentum separations between Weyl nodes, are taken into account in the recently proposed framework
of consistent hydrodynamics \cite{Gorbar:2017vph}. The latter includes several types of Chern--Simons
contributions in the electric current and charge densities that affect not only the electron transport in
Weyl semimetals \cite{Gorbar:2018vuh,Gorbar:2018sri}, but also their various collective excitations
\cite{Gorbar:2017vph,Sukhachov:2018nmg}. It is natural to ask, therefore, whether the hydrodynamic
regime could be also realized for the Fermi arc electrons and, if so, how it would affect the bulk
electron fluid.

A low sensitivity of the surface Fermi arc states to disorder makes them promising candidates
to sustain a \emph{surface electron fluid} in Weyl semimetals. If this is indeed the case, the
Fermi arcs could realize not only the ``Fermi level plumbing" \cite{Haldane:2014}, but act as true
aqueducts for the surface electron fluid. Because of the inevitable surface-bulk transitions and
interactions, of course, such a surface electron fluid should be necessarily coupled to the
bulk one.

In this study, we derive the hydrodynamic equations for the surface Fermi arc states
from the kinetic theory and phenomenologically describe the corresponding surface-bulk coupling.
Our principal finding is that the Fermi arc electron liquid modifies the boundary conditions for the bulk
one. Depending on the coupling parameters, the bulk flow in a slab of finite thickness could be
noticeably altered and even enhanced near the surface. In addition, we study the surface collective
modes in the hydrodynamic approximation. In agreement with the earlier studies
\cite{Song:2017,Andolina:2018,Losic:2018}, the presence of the Fermi arcs is manifested in a strongly
anisotropic dispersion relation of the surface plasmons. Additionally, we find that while the constant-frequency contours of the surface modes are given only by the elongated ellipses, the open hyperbolic
contours correspond to the bulk modes hybridized with the surface excitations similarly to the usual
semi-infinite plasma \cite{Barton:1979}. Finally, there is also a gapless surface mode, which is related
exclusively to the Fermi arcs. While such a mode resembles a conventional surface acoustic plasmon
\cite{Pitarke:2004}, which also has a linear dispersion relation, the Fermi arc mode is different and its
frequency is determined by the surface dispersion relation. These qualitative effects can be potentially used to experimentally verify the realization of the electron hydrodynamics in Weyl semimetals.

Our paper is organized as follows. In Sec.~\ref{sec:Model}, we introduce the phenomenological
hydrodynamic model of the Fermi arcs and discuss the coupling of the surface and bulk electron
fluids. The explicit realization of the coupling is given and the effects of the surface states on the
hydrodynamic flow are studied in Sec.~\ref{sec:hydro-flow}. Section~\ref{sec:SP} is devoted to the
investigation of the surface collective modes in the hydrodynamic approximation. Our results are
discussed and summarized in Sec.~\ref{sec:Summary}. Technical details, including the derivation
of the Fermi arc hydrodynamic equations and some auxiliary formulas are given in
Appendices~\ref{sec:App-derivation} and \ref{sec:App-ref}, respectively. Throughout this paper,
we set the Boltzmann constant $k_B=1$.

\section{Model}
\label{sec:Model}

In order to derive the hydrodynamic equations for the Fermi arc quasiparticles, we start from the
kinetic theory. As usual \cite{Landau:t10,Huang-book}, the Euler equation is obtained by calculating
the appropriate moments of the kinetic equation. In the presence of an electric field $\mathbf{E}$,
the latter reads as
\begin{equation}
\partial_t f^{\rm (FA)} -e\mathbf{E}\cdot \partial_{\mathbf{p}} f^{\rm (FA)} +\mathbf{v}_p^{\rm (FA)} \cdot\bm{\nabla} f^{\rm (FA)} =I^{\rm (FA)}_{\rm coll},
\label{Model-FA-KT-eq}
\end{equation}
where $-e$ is the electron charge, $\mathbf{p}=\left(p_x, p_z\right)$ is the two-dimensional
momentum of the surface quasiparticles, and $I^{\rm (FA)}_{\rm coll}$ denotes the collision
integral, whose effects on the Fermi arcs will be discussed later. In the hydrodynamic regime,
the distribution function describes local equilibrium, i.e.,
\begin{equation}
f^{\rm (FA)} =\delta(y-y_s)
\frac{1}{1+\rm{exp}\left(\frac{\epsilon_{p}^{\rm (FA)}-(\mathbf{u}^{\rm (FA)}\cdot\mathbf{p}) -\mu}{T}\right)},
\label{Model-FA-Fermi-Dirac}
\end{equation}
where $y_s$ denotes the surface coordinate, $s=\pm$ labels the bottom ($s=+$) and top ($s=-$) surfaces, respectively, $\mathbf{u}^{\rm (FA)}$ is the Fermi arc fluid velocity, $\mu$ is the electric chemical potential, and $T$ is temperature. For a slab of finite thickness $L_y$,
the coordinates of the top and bottom surfaces will be fixed at $y_{-}=L_y$ and $y_{+}=0$, respectively.
Here, for simplicity, we assume that the Fermi arcs are strongly localized at the sample's boundaries and,
therefore, the dependence of the distribution function on the transverse coordinate can be modeled by the
$\delta$-functions.

Further, we assume that the Weyl semimetal has a broken TR symmetry and contains two Weyl nodes separated by $2b$ along the $z$ direction in momentum space. Then, assuming a simple model (see, e.g., Ref.~\cite{Murakami:2014}), the dispersion relation for the Fermi arc states reads as
\begin{equation}
\label{model-FA-epsilon_p}
\epsilon_{p}^{\rm (FA)}= sv_Fp_x,
\end{equation}
where $v_F$ is the Fermi velocity. The linear energy dispersion implies that the Fermi arc quasiparticles have a constant velocity
\begin{equation}
\label{model-FA-v_p}
\mathbf{v}_p^{\rm (FA)}= \partial_\mathbf{p}\epsilon_{p}^{\rm (FA)} =sv_F\hat{\mathbf{x}}
\end{equation}
parallel to the $x$ axis. Therefore, it is reasonable to assume that their hydrodynamical velocity $\mathbf{u}^{\rm (FA)}$ also
points in the $x$ direction. In other words, there is no hydrodynamic flow due to
the Fermi arcs in the $z$ direction. Of course, the same is true for the surface electric current, which
can only flow along $\mathbf{v}_p^{\rm (FA)}$ [see Eq.~(\ref{derivation-FA-currents-J-FA}) in
Appendix~\ref{sec:App-derivation-Euler}]. We note that this is qualitatively different from the setup in
Ref.~\cite{Song:2017}, where a diffusive surface transport along the $z$ direction was allowed.

The derivation of the hydrodynamic equation for the Fermi arc electron fluid is given in
Appendix~\ref{sec:App-derivation}. In the inviscid limit, which might be justified in the
case of relatively small electron-electron collision times, the Euler equation for the surface
hydrodynamic velocity reads as
\begin{equation}
\label{model-Euler-FA}
\left(\partial_t+sv_F \partial_x\right) \frac{sw^{\rm (FA)}}{v_F}
\left(1+2\frac{u_x^{\rm (FA)}}{sv_F}\right) +en^{\rm (FA)} \left(1+\frac{u_x^{\rm (FA)}}{s v_F}\right) E_x  = I^{\rm (FA)}_s,
\end{equation}
where $I^{\rm (FA)}_s$ stems from the collision integral $I^{\rm (FA)}_{\rm coll}$ and describes
the surface-bulk transitions. The enthalpy and the fermion-number density of the Fermi arc states in equilibrium are derived in
Appendix~\ref{sec:App-derivation-Euler}. Their explicit expressions read as
\begin{eqnarray}
\label{model-w-FA-def}
w^{\rm (FA)} &=& \frac{b}{4\pi^2 v_F \hbar} \left(\mu^2 +\frac{\pi^2T^2}{3}\right),\\
\label{model-n-def}
n^{\rm (FA)} &=& \frac{\mu b}{2\pi^2v_F\hbar}.
\end{eqnarray}
Now, let us briefly discuss the bulk hydrodynamics. In the absence of external magnetic fields and vorticity,
the Navier--Stokes equation for the quasiparticles in a Weyl semimetal reads as \cite{Sachdev:2016,Gorbar:2018vuh}
\begin{equation}
\label{model-NS-B}
\partial_t \frac{w}{v_F^2} \mathbf{u} -\eta \Delta \mathbf{u} -\left(\zeta+\frac{\eta}{3}\right) \bm{\nabla}\left(\bm{\nabla}\cdot\mathbf{u}\right) +\bm{\nabla}P +en \mathbf{E} +\frac{w}{v_F^2\tau} \mathbf{u}
= \mathbf{I}_{\rm surf}.
\end{equation}
Here $w=\epsilon+P$ is the bulk enthalpy, $\epsilon$ is the bulk energy density, $P$ is the pressure, $\mathbf{u}$ is the bulk fluid velocity, $n$ is the bulk fermion-number density, $\eta$ and $\zeta$ are the shear and bulk dynamic viscosities, respectively. Note that, in the global equilibrium state without background electromagnetic fields and with vanishing fluid velocity, the energy density, the pressure, and the fermion-number density take the following explicit forms:
\begin{eqnarray}
\label{model-equilibrium-be}
\epsilon &=& \frac{15\mu^4+ 30\pi^2T^2 \mu^2 +7\pi^4T^4}{60\pi^2\hbar^3v_F^3},\\
P &=& \frac{\epsilon}{3},\\
n &=& \frac{\mu\left(\mu^2+\pi^2T^2\right)}{3\pi^2 \hbar^3 v_F^3}.
\label{model-equilibrium-ee}
\end{eqnarray}
In relativistic-like systems, the shear and bulk viscosities can be estimated as $\eta = w\tau_{ee}/4$ (see, e.g.,
Refs.~\cite{Alekseev:2016,Gooth:2017}) and $\zeta\approx 0$~\cite{Landau:t10}.
In our study, we use the electron-electron collision time $\tau_{ee}=\hbar/T$, which is consistent with the
experimental findings in Ref.~\cite{Gooth:2017}.

Compared to the conventional Navier--Stokes equation \cite{Landau:t6}, Eq.~(\ref{model-NS-B}) contains a few
additional contributions.
While the penultimate term on the left-hand side accounts for the charged nature of the electron fluid and describes the electrical force,
the term inversely proportional to the relaxation time $\tau$ is the hallmark feature of electron hydrodynamics in solids \cite{Gurzhi:1968}. It comes from the electron scattering on phonons and impurities and, as is clear from its explicit
dependence on the fluid velocity, breaks the Galilean invariance. As for the term on the right-hand side
of Eq.~(\ref{model-NS-B}), it describes the transfer of momentum between the surface and bulk fluids. By taking
into account that the Fermi arcs are localized on the surface, it can be modeled as follows:
\begin{equation}
\label{model-I-surf}
\mathbf{I}_{\rm surf}= -\sum_{s=\pm}\delta(y-y_s)I^{\rm (FA)}_s \hat{\mathbf{x}}.
\end{equation}
%where $y_{+}=0$ and $y_{-}=L_y$ are the coordinates of the two slab surfaces.
This inclusion of the source term in Eq.~(\ref{model-NS-B}) implies that the boundary conditions (BCs) for the electron fluid should be modified. Instead of the usual free-surface BCs, the fluid velocity and its derivatives should satisfy the following BCs at $y=y_s$:
\begin{equation}
\label{model-BC-ux-ys}
\eta \partial_{y}u_x(y_s) +\left(\zeta+\frac{\eta}{3}\right) \partial_x u_y(y_s) = s I^{\rm (FA)}_s.
\end{equation}
Additionally, the normal components of the electron fluid velocity should vanish on both surfaces, i.e.,
\begin{equation}
\label{model-BC-uy}
u_y(y_s) = 0.
\end{equation}
In order to illustrate the nontrivial effects of the Fermi arcs, we will also consider the benchmark case, where the
chiral shift is absent or directed normal to the surfaces of the slab. In such a simplified setup, there are no Fermi
arc surface states and the BCs for the bulk fluid velocity take the standard no-slip form, i.e.,
\begin{equation}
\label{model-BC-ux-no-slip}
u_x(y_s)=0
\end{equation}
or the free-surface form, i.e.,
\begin{equation}
\label{model-BC-ux-free-surface}
\partial_y u_x(y_s)=0.
\end{equation}
Since the Fermi arc fluid velocity affects only the $x$ component of the bulk flow, the $z$ component of the bulk fluid
velocity always satisfies the standard no-slip or free-surface BCs similar to those in Eqs.~(\ref{model-BC-ux-no-slip})
or (\ref{model-BC-ux-free-surface}). As we argue below, this benchmark case is useful to identify the
effects of the Fermi arcs on the hydrodynamic flow without making any \emph{a priori} assumptions about the state of the surface.

It is worth noting that the Navier-Stokes equation should be amended by the energy conservation relation as well as the
electric current continuity relation. While the latter has a profound effect on both the charge transport and collective
excitations, the former is important only for the thermoelectric effects and can usually be neglected in electron transport
\cite{Landau:t6}. In general, however, the effect of the energy conservation on the electron hydrodynamics may become
important when the fluid velocity is not small compared to the speed of sound $v_{sd}$ (note that $v_{sd}$ is close to
$v_F/\sqrt{3}$ in relativisticlike systems). Therefore, in what follows, we will assume that $|\mathbf{u}|\ll v_{sd}$ and the energy conservation relation can be ignored.

The electric current continuity relations for the surface and bulk states read as
\begin{eqnarray}
\label{model-divJ-FA-def}
\partial_t \rho^{\rm (FA)}_s + \bm{\nabla}_{\perp}\cdot\mathbf{J}^{\rm (FA)}_s &=& Q^{\rm (FA)}_s,\\
\label{model-divJ-def}
\partial_t \rho + \bm{\nabla}\cdot\mathbf{J} &=& -\sum_{s=\pm} \delta\left(y-y_s\right) Q^{\rm (FA)}_s,
\end{eqnarray}
respectively. Here $Q^{\rm (FA)}_s$ describes the electric charge transfer between the bulk and surface
states of the semimetal. Further, $\rho^{\rm (FA)}_s =-en^{\rm (FA)}\left(1+su_x^{\rm (FA)}/v_F\right)$ is the surface electric charge density and $\rho=-en$ is the bulk one. The surface and bulk electric currents are given by
\begin{eqnarray}
\label{model-Js-FA-def}
\mathbf{J}^{\rm (FA)}_s &=& sv_F \rho^{\rm (FA)}_s \hat{\mathbf{x}},\\
\label{model-J-def}
\mathbf{J} &=& -en\mathbf{u}+\sigma\mathbf{E} -\frac{e^2\left[\mathbf{b}\times\mathbf{E}\right]}{2\pi^2\hbar},
\end{eqnarray}
respectively. The explicit expressions for the Fermi arc charge and current densities are derived in Appendix~\ref{sec:App-derivation-Euler}.
Note that the expression for the bulk current (\ref{model-J-def})
includes the intrinsic conductivity $\sigma$, which was discussed in the holographic approach in
Refs.~\cite{Kovtun:2008kx,Hartnoll:2014lpa,Landsteiner:2014vua,Davison:2015taa,Lucas:2015lna}.
While $\sigma$ is important for the correct description of the normal flow in the presence of a nonzero chiral shift
\cite{Gorbar:2018vuh}, it plays no role in the longitudinal flow. The last term in Eq.~(\ref{model-J-def})
corresponds to the anomalous Hall effect~\cite{Ran,Burkov:2011ene,Burkov-Hook-Balents:2011,Grushin-AHE,Zyuzin-Burkov:2012,Goswami,Burkov-AHE:2014},
where $\mathbf{b}=b\hat{\mathbf{z}}$ is the chiral shift.

By integrating Eq.~(\ref{model-divJ-def}) in the vicinity of the surfaces, we obtain the following boundary condition for the normal component
of the bulk current:
\begin{equation}
\label{model-J-BC-y}
J_y(y_s) = -s Q^{\rm (FA)}_s.
\end{equation}
Formally, this implies that, because of the transitions between the surface and bulk states, the normal component of
the bulk current does not vanish on the surface.

Before concluding this section, it is instructive to sum up the general features of the surface and bulk flows,
as well as to reiterate the critical role that the BCs play in their interplay.
The surface $u_x^{\rm (FA)}$ and bulk $\mathbf{u}$  fluid velocities
are determined by the hydrodynamic equations~(\ref{model-Euler-FA}) and (\ref{model-NS-B}), respectively.
The bulk equation should be also supplemented by the appropriate BCs for the normal component of the fluid velocity, see
Eq.~(\ref{model-BC-uy}), as well as the BCs for the tangential components, see Eqs.~(\ref{model-BC-ux-ys})
or, in the absence of the Fermi arcs, Eqs.~(\ref{model-BC-ux-no-slip}) and (\ref{model-BC-ux-free-surface}). In addition, as we will see below,
the study of the longitudinal flow requires specifying either the Fermi arc fluid velocity at some contacts or
an explicit form of the transfer terms. In our study, we use the latter option that allows for a
self-consistent determination of the surface fluid velocity as well as the bulk flow.
As for the collective excitations, the use of the continuity relations (\ref{model-divJ-FA-def}) and (\ref{model-divJ-def}) will be
needed in order to determine the evolution of the electric charge. In this connection, it should be emphasized
that, because of the presence of the surface states, the normal component of the bulk electric current
(\ref{model-J-BC-y}) does not, in general, vanish at the boundary.

\section{Hydrodynamic flow}
\label{sec:hydro-flow}

In this section, we investigate a steady hydrodynamic flow in a slab of finite width in the $y$ direction and infinite in the $x$ and $z$ directions. We assume that the slab is sufficiently thick so that the interaction between the Fermi arcs on the opposite surfaces $y_{+}=0$ and $y_{-}=L_y$
is negligible and the arcs could be considered as independent. At the same time, the thickness
should be small enough in order for the surface flow to have a noticeable effect on the net
hydrodynamic flow. Without loss of generality, we also assume that a uniform
background electric field is applied in the $x$ direction. Note that this is the same setup that we used in Ref.~\cite{Gorbar:2018vuh}, where, however, the effects of the Fermi arcs were not taken into account.

Since the bulk electron fluid couples to the surface states, the transfer term on the right-hand side in
Eq.~(\ref{model-Euler-FA}) plays an important role in the hydrodynamic flow. We model it as follows:
\begin{equation}
\label{model-Icoll}
I^{\rm (FA)}_{s} = -\frac{w^{\rm (FA)} u_x^{\rm (FA)}}{v_F^2\tau_{sb}} + \frac{\alpha w u_x(y_s)}{v_F}.
\end{equation}
Here, the first term describes the transitions from the surface to the bulk with the rate determined by the relaxation time $\tau_{sb}$. The second term corresponds to the inflow from the bulk to the surface, where the rate is parametrized by a small numerical coefficient $\alpha$. These terms are derived by using the relaxation time approximation in Appendix~\ref{sec:App-transfer-term} [see Eqs.~(\ref{FA-bulk-connection-FA-to-Bulk}) and (\ref{FA-bulk-connection-Bulk-to-FA})]. For the surface to bulk transfer term, the relaxation time approximation might be indeed physical because the dissipation of the Fermi arc states is primarily due to the surface to bulk scatterings~\cite{Gorbar:2016aov}. In the hydrodynamic picture, this corresponds to the outflow into the bulk. As for the bulk to surface transfer term, it is estimated in a similar way. In general, the transfer terms in Eq.~(\ref{model-Icoll}) can be viewed as the leading terms in the gradient expansion about the global equilibrium state.

Taking into account that the right-hand sides of the bulk Navier-Stokes (\ref{model-NS-B}) and continuity
(\ref{model-divJ-def}) equations are nonzero only at the surfaces of the slab, it is reasonable to take
them into account only via the BCs. In particular, we will use Eq.~(\ref{model-BC-ux-ys})
for the velocity on the surface and Eq.~(\ref{model-J-BC-y}) for the normal component of the current.
Since the latter is not important for the longitudinal hydrodynamic transport, there is no need to specify
the explicit form of the transfer term $Q_s^{\rm (FA)}$. Therefore, the steady longitudinal flow in the bulk
is described by the following equation:
\begin{equation}
\label{hydro-flow-NS-B-1}
\eta \partial_{y}^2 u_x(y) -enE_x -\frac{w}{v_F^2\tau} u_x(y) =0.
\end{equation}
Here, as in Ref.~\cite{Gorbar:2018vuh}, we omitted $\bm{\nabla}P$ in the flow equation.
In view of the slab's geometry, there is no dependence on $x$ and $z$.
Then, the general solution to Eq.~(\ref{hydro-flow-NS-B-1}) reads as
\begin{equation}
\label{hydro-flow-NS-B-Sol}
u_x(y) = C_1e^{\lambda_x y} +C_2e^{-\lambda_x y} - \frac{en v_F^2 \tau E_x}{w},
\end{equation}
where $\lambda_x = \sqrt{w/(\eta v_F^2 \tau)}$ and the constants $C_1$ and $C_2$ are determined by the BCs. In particular, Eq.~(\ref{model-BC-ux-ys}) takes the form
\begin{equation}
\label{hydro-flow-BC-ux-ys}
s\eta \partial_y u_x(y_s) =  u_x(y_s) \frac{\alpha w}{v_F}\left[1 -\frac{w^{\rm (FA)}}{v_F\tau_{sb}} \left(sen^{\rm (FA)}E_x +\frac{w^{\rm (FA)}}{v_F\tau_{sb}}\right)^{-1} \right] +\frac{en^{\rm (FA)} E_x w^{\rm (FA)}}{v_F\tau_{sb}}\left(sen^{\rm (FA)}E_x +\frac{w^{\rm (FA)}}{v_F\tau_{sb}}\right)^{-1},
\end{equation}
where we used the following expression for $u_x^{\rm (FA)}$ obtained from Eqs.~(\ref{model-Euler-FA}) and (\ref{model-Icoll}):
\begin{equation}
\label{hydro-flow-ux-FA}
u_x^{\rm (FA)}(y_s) = -\left[ v_Fen^{\rm (FA)} E_x -\alpha w u_x(y_s)\right] \left(sen^{\rm (FA)}E_x +\frac{w^{\rm (FA)}}{v_F\tau_{sb}}\right)^{-1}.
\end{equation}
It is instructive to consider the following two limiting cases: (i) no transfer of electrons from the surface to the bulk $\tau_{sb} \to \infty$ and (ii) a very strong outflow from the surface to the bulk $\tau_{sb} \to 0$.

In the first case (i.e., $\tau_{sb} \to \infty$), we have
\begin{eqnarray}
\label{hydro-flow-tauFAB-to-infty-uFA}
u_x^{\rm (FA)}(y_s) &=& -sv_F +s\frac{\alpha w u_x(y_s)}{en^{\rm (FA)} E_x},\\
\label{hydro-flow-tauFAB-to-infty-BC-y=0}
s\eta \partial_{y}u_x(y_s) &=& \frac{\alpha w u_x(y_s)}{v_F}.
\end{eqnarray}
By noting that the velocity of the Fermi arcs might be as large as $v_F$, we could argue that the realization
of the hydrodynamic regime for the surface quasiparticles is unlikely at large $\tau_{sb}$. Also, in the limit $\alpha\to0$, i.e., when the Fermi arcs completely decouple, the BCs in Eq.~(\ref{hydro-flow-tauFAB-to-infty-BC-y=0})
reduce to the usual free-surface ones and $u_x^{\rm (FA)}(y_s)=-sv_F$.

In the opposite limit, i.e., $\tau_{sb} \to 0$, we obtain
\begin{eqnarray}
\label{hydro-flow-tauFAB-to-0-uFA}
u_x^{\rm (FA)} &=& 0,\\
\label{hydro-flow-tauFAB-to-0-BC-y=0}
s\eta \partial_{y}u_x(y_s) &=& en^{\rm (FA)}E_x.
\end{eqnarray}
In this case, there is a strong coupling between the surface and bulk fluids leading to the vanishing Fermi arc fluid velocity.
In addition, the boundary conditions for the bulk fluid are modified significantly and are affected by the electric field and the chiral
shift.

In a general case, the expression for the fluid velocity in the bulk of a Weyl semimetal can be obtained in an analytical
form by using the general solution in Eq.~(\ref{hydro-flow-NS-B-Sol}) and the BC in Eq.~(\ref{hydro-flow-BC-ux-ys}).
The corresponding expression is, however, rather cumbersome and not very informative. Therefore, instead of presenting it
here, we illustrate the key features of the flow, as well as the nontrivial effects of the Fermi arcs by using a representative set of model parameters. In particular, we use the following material constants:
\begin{equation}
\label{hydro-flow-results-material}
v_F= 1.4\times10^7~\mbox{cm/s}, \quad b = 3~\mbox{nm}^{-1}, \quad \tau= 3\times 10^{-10}~\mbox{s}, \quad \varepsilon_{e}= 13,
\end{equation}
which are comparable to those in Refs.~\cite{Autes-Soluyanov:2016,Gooth:2017,Kumar-Felser:2017,Razzoli-Felser:2018}.
(Note that the estimate for the electric permittivity $ \varepsilon_{e}$ is based on the dielectric constants of tungsten
\cite{Ordal:1988} and phosphorus \cite{Nagahama:1985}.)
By default, we also use the following values of other model parameters:
\begin{equation}
\label{hydro-flow-results-default}
\mu=25~\mbox{meV}, \quad T=10~\mbox{K}, \quad E_x=1~\mbox{V/m}, \quad L_{y,0}=10~\mu\mbox{m}, \quad \tau_{sb}=1~\mbox{ns},
\quad \alpha=10^{-4}.
\end{equation}

\begin{figure*}[!ht]
\begin{center}
\includegraphics[width=0.45\textwidth]{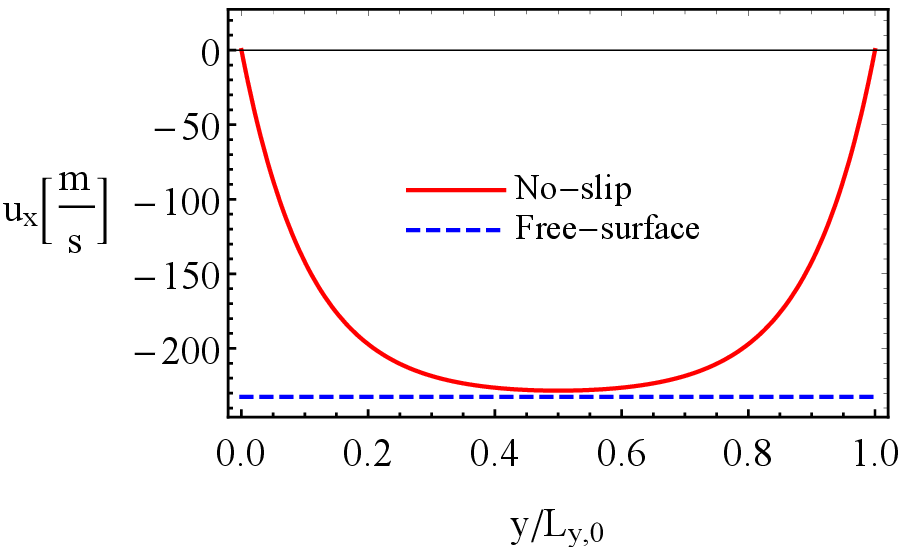}\hfill
\includegraphics[width=0.45\textwidth]{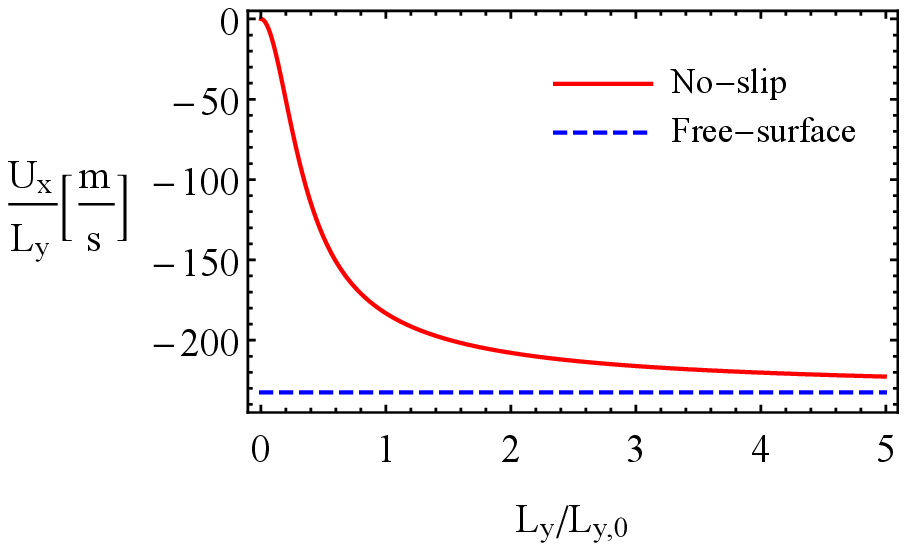}
\end{center}
\caption{The longitudinal flow velocity $u_x$ as a function of $y$ (left panel) and the longitudinal
flow velocity integrated over the channel width, $U_x=\int_0^{L_y}u_x(y) dy$, as a function of the
slab width $L_y$ (right panel). Red solid and blue dashed lines correspond to the standard no-slip
(\ref{model-BC-ux-no-slip}) and free-surface (\ref{model-BC-ux-free-surface}) BCs, respectively.}
\label{fig:hydro-flow-results-no-FA-ux}
\end{figure*}

In order to better clarify the role of the surface states, we start with the benchmark case without the Fermi arcs on the surfaces of the slab. Such a situation is realized naturally when the chiral
shift is absent (e.g., in Dirac semimetals) or when its direction
is perpendicular to the surfaces.
In this special case, the bulk fluid velocity in a Weyl semimetal slab is given by \cite{Gorbar:2018vuh}
\begin{equation}
\label{FA-bulk-connection-u-no-FA}
u_x(y) = -\frac{v_F^2\tau enE_x}{w} \left(1-\delta \frac{\cosh{\left(\lambda_x y - \lambda_x L_y/2\right)}}{\cosh{\left(\lambda_x L_y/2\right)}} \right),
\end{equation}
where $\delta=1$ and $0$ correspond to the standard no-slip and free-surface BCs given in
Eqs.~(\ref{model-BC-ux-no-slip}) and (\ref{model-BC-ux-free-surface}), respectively. The corresponding
profile of the longitudinal bulk flow velocity as a function of the $y$ coordinate is shown in the left
panel of Fig.~\ref{fig:hydro-flow-results-no-FA-ux}. Additionally, in the right panel of
Fig.~\ref{fig:hydro-flow-results-no-FA-ux}, we show the dependence of the flow velocity integrated
over the channel width, i.e., $U_x=\int_0^{L_y}u_x(y) dy$, on the slab width $L_y$.

%%%%%%%%%%%%%%%%%%
%%%%%%%%%%%%%%%%%%

\begin{figure*}[!ht]
\begin{center}
%\hspace{-0.32\textwidth}(a)\hspace{0.32\textwidth}(b)\hspace{0.32\textwidth}(c)\\[0pt]
\includegraphics[width=0.45\textwidth]{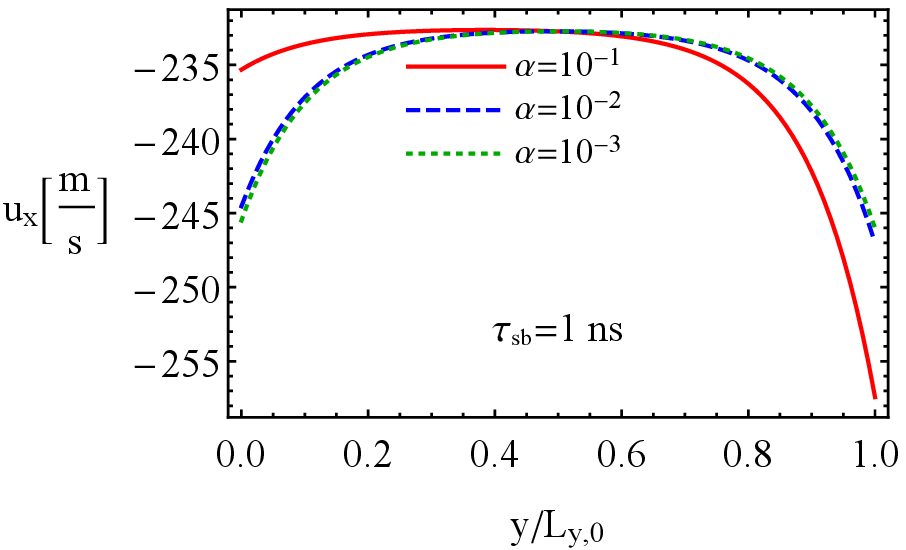}\hfill
\includegraphics[width=0.45\textwidth]{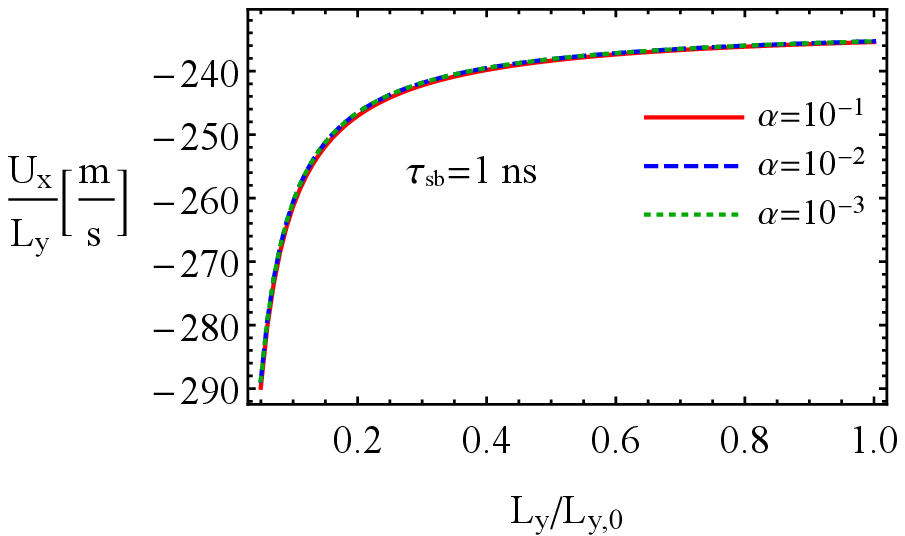}
\end{center}
\caption{The longitudinal flow velocity $u_x$ as a function of the $y$ coordinate (left panel) and the
longitudinal flow velocity integrated over the channel width, $U_x=\int_0^{L_y}u_x(y) dy$, as a function
of the slab width $L_y$ (right panel) for a few fixed values of $\alpha $. To obtain the numerical results,
we used the BC in Eq.~(\ref{hydro-flow-BC-ux-ys}) and set $\tau_{sb}=1~\mbox{ns}$.}
\label{fig:hydro-flow-results-ux-alpha}
\end{figure*}

Now, let us discuss the case of hydrodynamic flow with Fermi arcs on the surfaces.
The longitudinal flow velocity $u_x$ and the flow velocity
integrated over the channel width $U_x$ for a few values of the coupling parameter $\alpha$ are shown in the left and right panels of
Fig.~\ref{fig:hydro-flow-results-ux-alpha}, respectively. As we see, the presence of the Fermi arcs
enhances the fluid velocity near the boundaries when $\tau_{sb}$ is sufficiently small and the
transitions from the bulk to the surface, quantified by $\alpha $, are weak. This is in a drastic
contrast to the case of the conventional no-slip or free-surface BCs presented in
Fig.~\ref{fig:hydro-flow-results-no-FA-ux}. The increase of both fluid velocity and integrated fluid flow velocity is caused by the Fermi arc fluid that tends to push the bulk one near the surfaces. As expected, the net enhancement of the flow is noticeable only for sufficiently small
widths of the slab. It is interesting to note that the fluid velocity profile is rather sensitive to the
value of $\alpha$, which parametrizes the rate of transitions from the bulk to the surface. At the same time, the dependence of $U_x$ on $\alpha $ is very weak. We checked that, with increasing
$\tau_{sb}$ and/or $\alpha $, the effect of the Fermi arcs on the bulk hydrodynamic flow weakens
and gradually changes to a suppression of the flow velocity near the surfaces. It should be noted that, as expected on the basis of Eq.~(\ref{hydro-flow-tauFAB-to-infty-uFA}), the Fermi arc fluid velocity also grows with $\tau_{sb}$ and could eventually reach large enough values so that the hydrodynamic approach for the surface states becomes
unreliable. Therefore, $\tau_{sb}$ should remain sufficiently small to allow for the Fermi arcs hydrodynamics.

%%%%%%%%%%%%%%%%%%
%%%%%%%%%%%%%%%%%%

\section{Surface collective modes}
\label{sec:SP}

In this section, we study the effect of the Fermi arcs on the collective modes in a semi-infinite Weyl
semimetal in the hydrodynamic regime. In particular, we focus on the surface plasmons. We assume that the Weyl semimetal is located in the upper half-space ($y>0$) and the vacuum in the lower half ($y<0$). Therefore, in the notation of Sec.~\ref{sec:Model}, $s=+$. Henceforth, for simplicity, we omit the corresponding subscript in the Fermi arc variables.

Previously, the Fermi arc plasmons have been already studied in
Refs.~\cite{Song:2017,Andolina:2018,Losic:2018}. While Ref.~\cite{Song:2017} employs a simple
phenomenological model, the authors of \cite{Andolina:2018} provide a more rigorous quantum-mechanical
non-local approach. In this study, by following the approach similar to that used in
Refs.~\cite{Ritchie:1969,Barton:1979,Pitarke:2006}, we employ the hydrodynamic approximation
without the retardation effects in Maxwell's equations. In this case, the oscillating electric potential $\phi(t,\mathbf{r})$ is governed  by Poisson's equation
\begin{equation}
\label{SP-Poisson}
\Delta \phi(t,\mathbf{r})=\frac{4\pi e}{\varepsilon_e(y)} \delta n(t,\mathbf{r}),
\end{equation}
where $\delta n(t,\mathbf{r})$ describes
the deviations of the bulk fermion-number density from its equilibrium value,
$\varepsilon_{e}(y)=\theta(y)\varepsilon_e +\theta(-y)$ is the electric permittivity of the
system, and $\theta(y)$ is the Heaviside step function. The omission of the retardation
effects is formally equivalent to setting $c\to\infty$, which implies that the effects of
oscillating magnetic fields can be neglected as well.

By taking into account that the sought collective modes are localized on the surface of a
Weyl semimetal, we look for the solutions in the form of plane waves, $\delta X(t,\mathbf{r})
= \delta X(y)\, e^{-i\omega t+i\mathbf{k}_{\perp}\mathbf{r}_{\perp}}$, where $\omega$ is the
frequency, $\mathbf{k}_{\perp}=\left\{k_x,0,k_z\right\}$ is the surface wave vector,
and $\delta X$ is an oscillating hydrodynamic variable, e.g., $\delta \mu$, whose amplitude
may depend on the $y$ coordinate.

Following the arguments in Refs.~\cite{Ritchie:1969,Barton:1979,Pitarke:2006}, we can neglect
the effects of the energy conservation relation and set $\delta T=0$. Then, all oscillating
thermodynamical variables can be expressed in terms of the bulk fermion-number density
\begin{equation}
\label{SP-tilde-all}
\delta P = \tilde{P} \delta n, \quad \delta w = \tilde{w} \delta n, \quad \delta w^{\rm (FA)} = \tilde{w}^{\rm (FA)} \delta n^{\rm (FA)},
\end{equation}
where
\begin{eqnarray}
\label{SP-tP-exp}
\tilde{P}&=&\frac{\partial P}{\partial \mu} \left(\frac{\partial n}{\partial \mu}\right)^{-1} = \mu \frac{\mu^2+\pi^2T^2}{3\mu^2+\pi^2T^2} \stackrel{T\to0}{=} \frac{\mu}{3},\\
\label{SP-tw-exp}
\tilde{w} &=& \frac{\partial w}{\partial \mu} \left(\frac{\partial n}{\partial \mu}\right)^{-1} = 4\tilde{P}, \\
\label{SP-twFA-exp}
\tilde{w}^{\rm (FA)} &=&  \frac{\partial w^{\rm (FA)}}{\partial \mu} \left(\frac{\partial n^{\rm (FA)}}{\partial \mu}\right)^{-1} = \mu.
\end{eqnarray}
Furthermore, by assuming a gradient flow, which is consistent with the omission of vorticity,
the oscillations of the flow velocity can be expressed in terms of the velocity potential $\psi(t,\mathbf{r})$ as follows:
\begin{equation}
\label{SP-psi-def}
\delta\mathbf{u}(t,\mathbf{r}) = -\bm{\nabla}\psi(t,\mathbf{r}).
\end{equation}
Then, the Navier-Stokes equation (\ref{model-NS-B}), where the surface-bulk transitions are accounted for by the BCs, takes the form
\begin{equation}
\label{SP-NS-1}
i\omega\frac{w}{v_F^2} \bm{\nabla} \psi + \tilde{P} \bm{\nabla} \delta n +\left(\zeta + \frac{4}{3}\eta\right) \Delta \bm{\nabla} \psi - en \bm{\nabla} \phi - \frac{w}{v_F^2 \tau} \bm{\nabla}\psi=0.
\end{equation}
In order to obtain the solution for $\psi(y)$, we will reexpress $\delta n$ and $\phi$ in terms of the velocity potential.
By using the continuity relation (\ref{model-divJ-def}), the oscillating electric charge density reads as
\begin{equation}
\label{SP-J-conserv-eq}
\delta n = \frac{i n}{\omega} \Delta \psi -\frac{i\sigma}{\omega e} \Delta \phi.
\end{equation}
This implies that the Poisson equation (\ref{SP-Poisson}) inside the semimetal takes the form
\begin{equation}
\label{SP-Poisson-2}
\Delta \phi=i\frac{4\pi e n}{\varepsilon_e\omega \tilde{\sigma}} \Delta \psi,
\end{equation}
where $\tilde{\sigma}=\left[1+i 4\pi \sigma/(\varepsilon_e \omega)\right]$.
By making use of the last two equations, we can rewrite the Navier--Stokes equation (\ref{SP-NS-1}) as
\begin{equation}
\label{SP-NS-3}
\Delta \left[\omega^2\frac{w}{v_F^2} +\frac{\tilde{P} n}{\tilde{\sigma}} \Delta -i\omega \left(\zeta + \frac{4}{3}\eta\right)\Delta -\frac{4\pi e^2 n^2}{\varepsilon_e \tilde{\sigma}} +i\omega \frac{w}{v_F^2 \tau}\right] \psi=0.
\end{equation}
As is clear, the solution for $\psi$ that decreases in the bulk of the semimetal has the following form:
\begin{equation}
\label{SP-NS-sol}
\psi(y) = C_0^{\psi} e^{-k_{\perp}y} + \sum_{j=\pm}C_j^{\psi} e^{-\lambda_j y},
\end{equation}
where
\begin{equation}
\label{SP-example-lambda-eta}
\lambda_{\pm} = \pm\frac{\sqrt{3\omega_p^2  + k_{\perp}^2\left(3K^2 -4iv_F^2\eta \tilde{\sigma} \omega/w\right) -3\omega \tilde{\sigma} \left(\omega +i/\tau\right)}}{\sqrt{3K^2 -4i\eta v_F^2 \tilde{\sigma} \omega/w}}.
\end{equation}
Here, we used the shorthand notation $K^2=v_F^2n\tilde{P}/w$, which approaches $v_F^2/3$ as $T \to 0$, and introduced the plasma frequency $\omega_{p}^2=4\pi e^2v_F^2 n^2/(\varepsilon_e w)$. Note that when $\lambda_{\pm}$ are purely imaginary, the corresponding modes are hybridized surface-bulk excitations.
When the bulk viscosity, dissipation, and the intrinsic conductivity are ignored, the
expression in Eq.~(\ref{SP-example-lambda-eta}) reads as
\begin{equation}
\label{SP-example-lambda}
\lambda^{(0)}_{\pm} = \pm\frac{\sqrt{\omega_p^2 +K^2k_{\perp}^2 -\omega^2}}{K}.
\end{equation}
As is clear, only $\lambda^{(0)}_{+}$ corresponds to a mode localized on the surface.

Having determined $\psi(y)$, we can now find the expressions for the electric potentials both inside $y>0$ and
outside $y<0$ the semimetal
\begin{eqnarray}
\label{SP-phiSM}
\phi^{y>0}(y) &=& C^{\phi} e^{-k_{\perp}y} +i\frac{4\pi e n}{\varepsilon_e \omega \tilde{\sigma}} \sum_{j=\pm }C_{j}^{\psi} e^{-\lambda_{j} y},\\
\label{SP-phiV}
\phi^{y<0}(y) &=& \tilde{C}^{\phi} e^{k_{\perp}y}.
\end{eqnarray}
Similarly to the fluid flow, the BCs are also important for the surface collective modes.
For the oscillating fluid velocity, we impose the same BCs as for the flow in Eqs.~(\ref{model-BC-ux-ys}) and (\ref{model-BC-uy}),
i.e.,
\begin{eqnarray}
\label{SP-BC-ux}
\eta \partial_{y}\delta u_x(0)&=& -ik_x \eta \partial_{y} \psi(0) = I^{\rm (FA)},\\
\label{SP-BC-uy}
\delta u_y(0)&=&-\partial_y\psi(0) =0.
\end{eqnarray}
As is clear from the above equations, the transfer term $I^{\rm (FA)}$ in Eq.~(\ref{SP-BC-ux}) should be set to zero. It is worth noting that, in a general case where the gradient approximation cannot be employed, $I^{\rm (FA)}\neq0$. In particular, an explicit form of the transfer term should be defined in order to solve the hydrodynamic equations for the longitudinal flow in Sec.~\ref{sec:hydro-flow}.

The BCs for the electric potential have the following standard form:
\begin{eqnarray}
\label{SP-BC-phi}
\phi^{y>0}(0) &=& \phi^{y<0}(0),\\
\label{SP-BC-dphi}
\varepsilon_e\partial_y\phi^{y>0}(0) -\partial_y\phi^{y<0}(0) &=& 4\pi e \delta n^{\rm (FA)},
\end{eqnarray}
where the normal component of the oscillating electric field $\delta\mathbf{E}(y)=-\bm{\nabla}\phi(y)$ has a jump connected
with the singular contribution of the Fermi arcs. According to Ref.~\cite{Barton:1979}, the bulk states themselves should
not induce a localized (singular) surface charge density.
The oscillations of the surface fermion
number density $\delta n^{\rm (FA)}$ can be obtained by using Eqs.~(\ref{model-divJ-FA-def}) and (\ref{model-J-BC-y}), i.e.,
\begin{equation}
\label{SP-dnFA}
\delta n^{\rm (FA)} = -\frac{v_Fk_x n^{\rm (FA)} e\phi^{y>0}(0)}{\mu(\omega-v_Fk_x)}
-i\frac{\sigma \partial_y \phi^{y>0}(0)}{e(\omega-v_Fk_x)}.
\end{equation}
As for the oscillating Fermi arc fluid velocity $\delta u_x^{\rm (FA)}$, it can be obtained from the surface Euler equation (\ref{model-Euler-FA}).

Another boundary condition can be derived from the $y$ component of Eq.~(\ref{SP-NS-1}) after expressing
$\Delta \psi$ in terms of $\delta n$ [see Eqs.~(\ref{SP-J-conserv-eq}) and (\ref{SP-Poisson-2})] and utilizing the BC in Eq.~(\ref{SP-BC-uy}). The explicit form of this new BC reads as
\begin{equation}
\label{SP-BC-add}
\left[\tilde{P}  -i\frac{\omega \tilde{\sigma}}{n}\left(\zeta + \frac{4}{3}\eta\right)
\right] \partial_y \delta n(0) = en \partial_y\phi^{y>0}(0).
\end{equation}
Equations (\ref{SP-BC-uy})--(\ref{SP-BC-add}) are sufficient to reexpress all integration constants in Eqs.~(\ref{SP-NS-sol}), (\ref{SP-phiSM}), and (\ref{SP-phiV}) in terms of a single constant
that is then fixed by a normalization condition. Also, after satisfying all the boundary conditions, one can determine the
dispersion relations for the surface modes. As for the dispersion relations for the hybridized surface-bulk modes,
they are obtained from Eq.~(\ref{SP-example-lambda-eta}), where $\lambda_{\pm}\to\lambda$ is a continuous variable \cite{Barton:1979}.

For the sake of simplicity, we neglect the effects of viscosity ($\eta\to0$), dissipation
($\tau\to\infty$), and intrinsic conductivity ($\sigma\to0$) in the rest of this section. It is also instructive to start from the
benchmark case without the Fermi arcs on the surface of a Weyl semimetal. Then, after satisfying all BCs, Eq.~(\ref{SP-BC-add}) gives the following relation for the modes localized on the surface:
\begin{equation}
\label{SP-example-no-FA-char-eq}
K^2 \lambda^{(0)}_{+} \left(\lambda^{(0)}_{+}+k_{\perp}\right) = \frac{\omega_p^2}{1+\varepsilon_e},
\end{equation}
where $K$ and $\omega_p$ are defined after Eq.~(\ref{SP-example-lambda-eta}). The corresponding positive solution is given by
\begin{equation}
\label{SP-no-FA-omega}
\omega = \frac{1}{\sqrt{2}}\left(\frac{2\varepsilon_e\omega_p^2}{1+\varepsilon_e} + K^2 k_{\perp}^2 + Kk_{\perp} \sqrt{\frac{4\omega_p^2}{1+\varepsilon_e}+K^2k_{\perp}^2}\right)^{1/2} \approx \frac{\omega_p \sqrt{\varepsilon_e}}{\sqrt{1+\varepsilon_e}} +\frac{K k_{\perp}}{2 \sqrt{\varepsilon_e}} +\frac{K^2 k_{\perp}^2 \sqrt{1+\varepsilon_e} (2\varepsilon_e-1)}{8 \omega_p \varepsilon_e^{3/2}} + O(k_{\perp}^3).
\end{equation}
Note that the long-wavelength approximation is well defined and consistent with the nonretarded regime only at $\omega<ck_{\perp}$. By taking into account the large value of $c$, the range of validity of the above result extends to rather
small values of the wave vector, $k_{\perp}\simeq \omega_p/c$. The surface plasmon frequency $\omega$ in
Eq.~(\ref{SP-no-FA-omega}) qualitatively agrees with the results obtained in
Refs.~\cite{Ritchie:1969,Barton:1979,Pitarke:2006}. As we see, the spectrum of the surface plasmons
is isotropic and has a nonzero gap. Moreover, the value of the gap agrees with that obtained in
Ref.~\cite{Ritchie:1957} after setting $\varepsilon_e=1$.

\begin{figure*}[!ht]
\begin{center}
\includegraphics[width=0.45\textwidth]{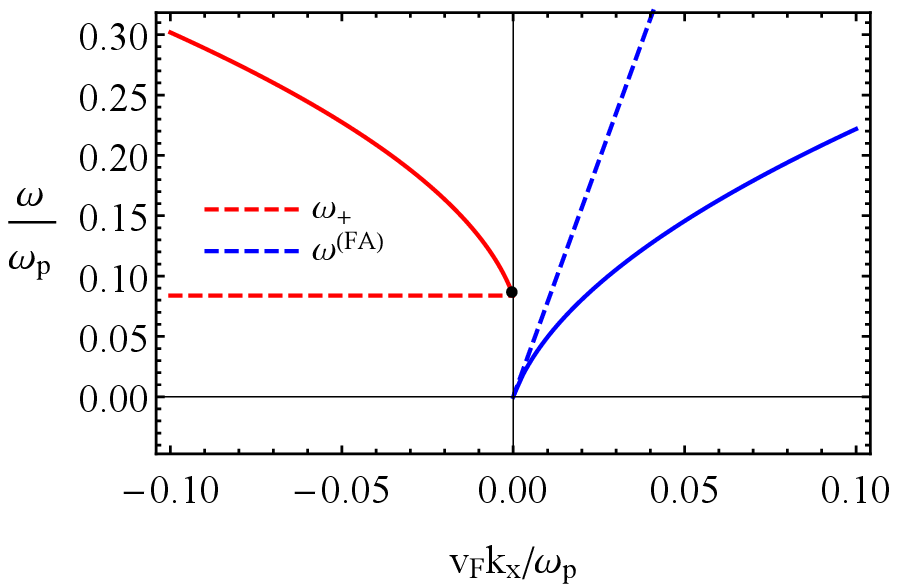}\hfill
\includegraphics[width=0.45\textwidth]{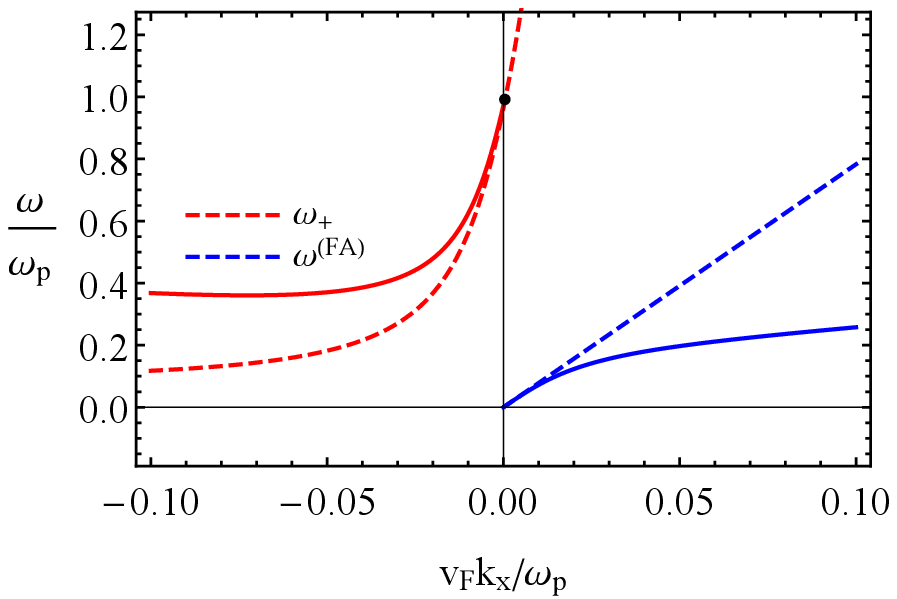}
\end{center}
\caption{The solutions for the surface collective modes in the presence of the Fermi arcs. Solid and dashed
lines correspond to the exact solutions of the characteristic equation and the approximate ones in
Eqs.~(\ref{SP-FA-omega-pm}) and (\ref{SP-FA-omega-3}), respectively. The red lines correspond to the gapped
surface plasmons and the blue lines describe the Fermi arc surface mode. Left and right panels show the
results for $k_z=0$ and $k_z=0.1\,\omega_p/v_F$. Black dots indicate the frequencies at which the surface
modes hybridize with the bulk ones.
}
\label{fig:SP-FA-m=m0}
\end{figure*}

\begin{figure*}[!ht]
\begin{center}
%\hspace{-0.32\textwidth}(a)\hspace{0.32\textwidth}(b)\hspace{0.32\textwidth}(c)\\[0pt]
\includegraphics[width=0.45\textwidth]{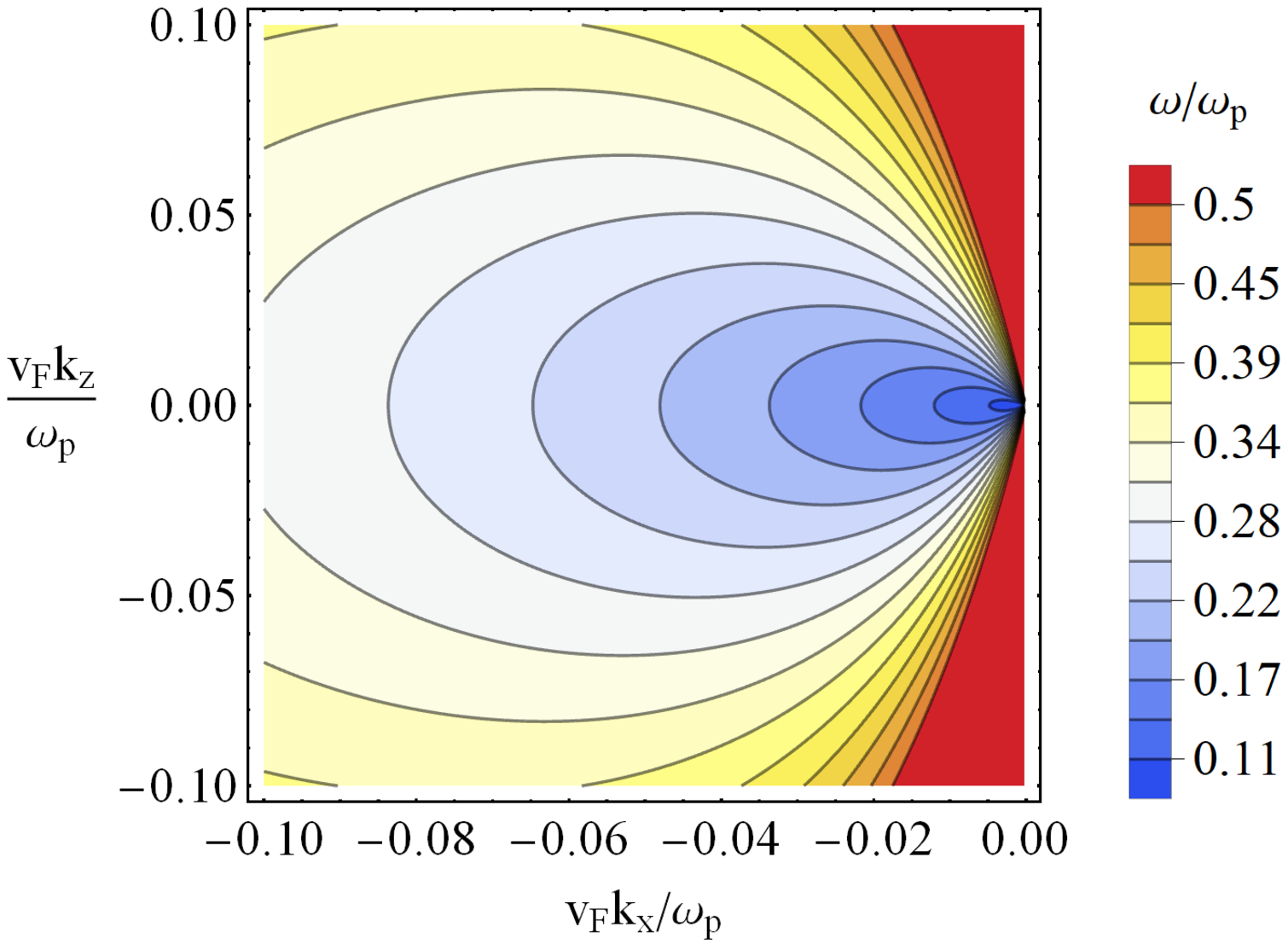}\hfill
\includegraphics[width=0.45\textwidth]{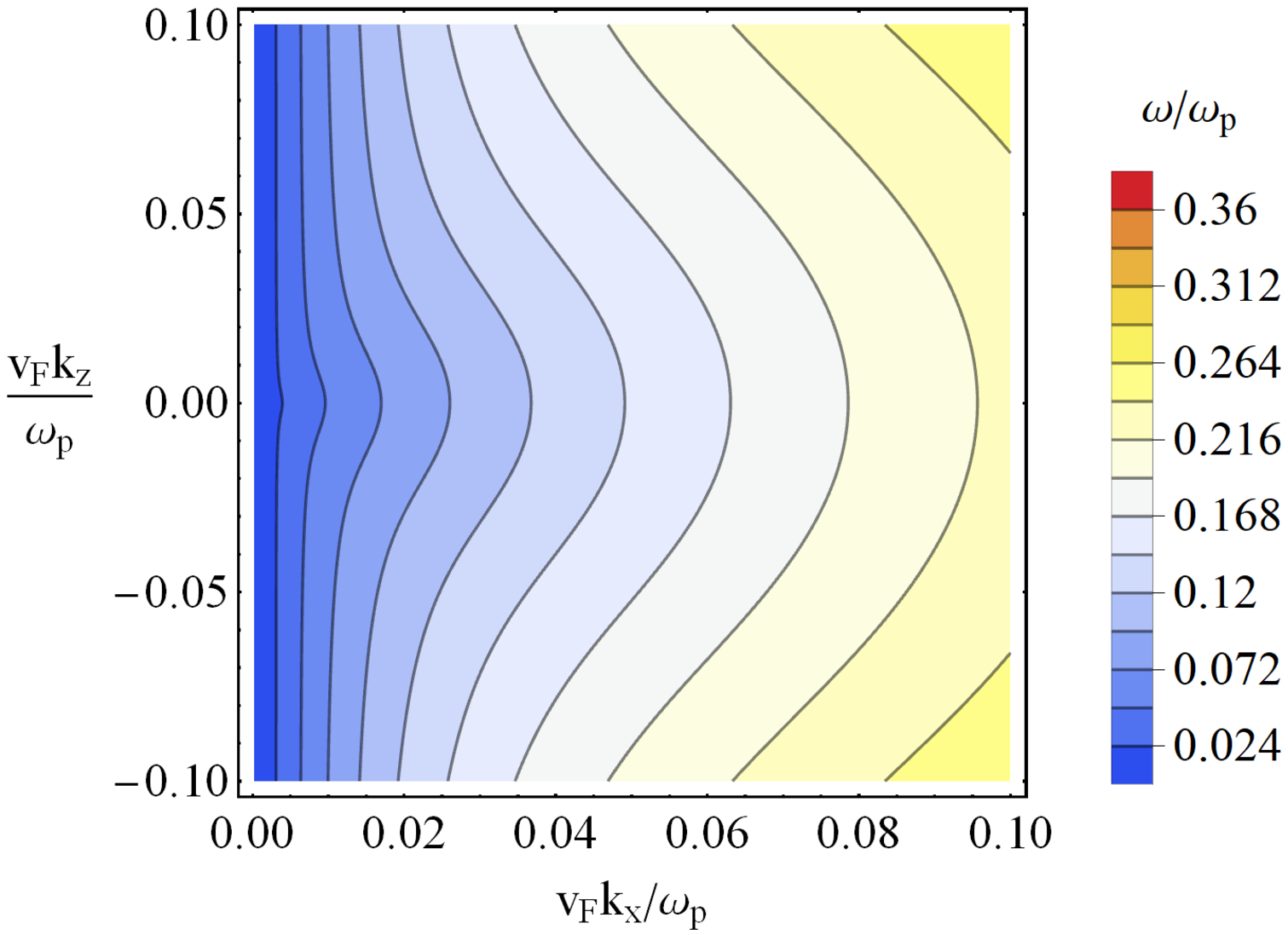}
\end{center}
\caption{The contour plots for the surface plasmon $\omega_{+}$ (left panel) and the
Fermi arc mode $\omega^{\rm (FA)}$ (right panel) frequencies. The group velocity is normal to the contour lines and only the results for $\omega>0$ are presented.
}
\label{fig:SP-FA-contour-num-m=m0}
\end{figure*}

Now let us analyze the effect of the Fermi arcs on the surface collective modes. Because of a nonzero
surface charge density in Eq.~(\ref{SP-BC-dphi}), the characteristic equation becomes more complicated
\begin{equation}
\label{SP-FA-char-eq}
K^2 \lambda^{(0)}_{+} (k_{\perp}+\lambda^{(0)}_{+}) = \frac{\omega_p^2 \left[\omega_b -(\omega-v_Fk_x)\right]}{\omega_b -(1+\varepsilon_e)(\omega-v_Fk_x)},
\end{equation}
where $\omega_b = 2e^2b k_x/(\pi \hbar k_{\perp})$. By solving the characteristic equation in the long-wavelength approximation, we obtain the following dispersion relations:
\begin{eqnarray}
\label{SP-FA-omega-pm}
\omega_{\pm} &=& \frac{1}{2(1+\varepsilon_e)}\left( \omega_b \pm \sqrt{\omega_b^2 +4\varepsilon_e(1+\varepsilon_e) \omega_p^2} \right) +O(k_{\perp}),\\
\label{SP-FA-omega-3}
\omega^{\rm (FA)} &=& v_F k_x +\frac{k_{\perp} K\omega_b}{\varepsilon_e \omega_p} +O(k_{\perp}^2).
\end{eqnarray}
As in the simplified case without Fermi arcs, the long-wavelength approximation is well-defined and consistent with the nonretarded regime for $k_{\perp}\gtrsim \omega/c$, which is obviously the case for $\omega^{\rm (FA)}$ in Eq.~(\ref{SP-FA-omega-3}). In addition, as in usual metals (see, e.g., Ref.~\cite{Pitarke:2006} and references therein), the hydrodynamic approximation is applicable even for the phase velocities of order $v_F$. It is worth noting, however, that the Landau damping, which is usually not captured in the hydrodynamic approach, could become relevant when the phase velocity of the surface mode with the frequency $\omega^{\rm (FA)}$ is smaller than the quasiparticle velocity $v_F$ (see, e.g., Ref.~\cite{Landau:t10}). By making use of the approximate dispersion relation in Eq.~(\ref{SP-FA-omega-3}) (as well as the results in Figs.~\ref{fig:SP-FA-m=m0} and \ref{fig:SP-FA-contour-num-m=m0}), we checked that this is not the case here because $\omega^{\rm (FA)}/k_x> v_F$ at sufficiently small values of $k_x$. In general, however, the Landau damping could provide an additional dissipation mechanism and should be included in a more rigorous treatment beyond the hydrodynamic approximation.

The modes with $\omega_{\pm}$ can be identified with the surface plasmons. Their frequencies are similar to those in Eq.~(16) of Ref.~\cite{Andolina:2018}. It can be also verified that, in agreement with the analysis in
Ref.~\cite{Song:2017}, the dispersion relations in Eq.~(\ref{SP-FA-omega-pm}) have discontinuities
$\propto\mbox{sign}(k_x)$ at $k_x=0$, namely $\left(\lim_{k_x\to+0} - \lim_{k_x\to-0}\right)\lim_{k_z\to0}
\omega_{\pm} = |\omega_b|/(1+\varepsilon_e)$. Such discontinuities disappear at $k_z\neq0$.
It is important to note that, depending on the chiral shift, $\omega_{+}$ ($\omega_{-}$) at $k_x>0$
($k_x<0$) could be significantly larger than $\omega_p^2 +K^2k_{\perp}^2$. Then, by taking into
account that the characteristic root defined in Eq.~(\ref{SP-example-lambda}) becomes purely imaginary,
the corresponding excitations should be identified with the hybridized surface-bulk modes
\cite{Barton:1979,Pitarke:2006} and, henceforth, will be omitted.

In contrast to the surface plasmons, which exist even in the absence of the Fermi arcs, the mode with
the dispersion relation in Eq.~(\ref{SP-FA-omega-3}) originates exclusively from the surface states.
It is somewhat reminiscent of the usual surface acoustic plasmon \cite{Pitarke:2004}
with a linear dispersion relation. However, the new mode stemming from the Fermi arcs has a rather
unconventional directional dependence.

The frequencies of the surface plasmons and the Fermi arc mode are presented in Fig.~\ref{fig:SP-FA-m=m0}, where the solid and dashed lines correspond to the numerical solutions of Eq.~(\ref{SP-FA-char-eq}) and the approximate
dispersion relations in Eqs.~(\ref{SP-FA-omega-pm}) and (\ref{SP-FA-omega-3}), respectively. Black dots
indicate the frequencies at which the surface modes hybridize with the bulk ones. In agreement
with the previous analysis in Ref.~\cite{Song:2017}, we found three roots of the characteristic equation
(\ref{SP-FA-char-eq}). Two of them correspond to $\omega_{\pm}$ in Eq.~(\ref{SP-FA-omega-pm}), which
are related by the transformation $\omega\to-\omega$ and $\mathbf{k}_{\perp}\to -\mathbf{k}_{\perp}$
(the modes with $\omega<0$ are not shown in the figures). The third solution describes a
gapless Fermi arc surface mode with the dispersion relation that, at leading order in small
$\mathbf{k}_\perp$, is approximately given by Eq.~(\ref{SP-FA-omega-3}).
In addition, as one can see from Fig.~\ref{fig:SP-FA-m=m0}, the leading order approximate expressions (\ref{SP-FA-omega-pm}) and (\ref{SP-FA-omega-3}) describe the collective modes rather well only when the wave vector is sufficiently small.

%%%%%%%%%%%%%%%%%%
%%%%%%%%%%%%%%%%%%

In order to clarify the dependence of the surface mode frequencies on the wave vectors, we present the
contour plots in momentum space for the positive frequency plasmon and the Fermi arc surface mode in
Fig.~\ref{fig:SP-FA-contour-num-m=m0}. As one can see from the left panel in Fig.~\ref{fig:SP-FA-contour-num-m=m0}, the contour lines of the gapped surface plasmon are closed ellipses elongated in the direction defined by the Fermi arcs dispersion relation, i.e., $k_x$. The contours of the Fermi arc mode are bell-shaped with the maximum at $k_z=0$.
%that, with increasing the momenta, gradually become parallel to $k_z$.

By noting that the group velocities of the surface collective modes are given by the derivatives of their
frequencies with respect to momenta, they can be represented by the vectors normal to the contour
lines. Then, as is clear from Fig.~\ref{fig:SP-FA-contour-num-m=m0}, while there is a preferred direction defined by $k_x$, the surface plasmons could also propagate radially similarly to the conventional surface plasmons with the frequency given in Eq.~(\ref{SP-no-FA-omega}). The gapless modes, on the other hand, always propagate in one direction, although there is a noticeable spreading, especially when the chiral shift is large.

It should be emphasized that the constant-frequency contours for the plasmon modes obtained in this study
are closed ellipses. This may appear to be qualitatively different from the open hyperbolic contours in
Refs.~\cite{Song:2017,Andolina:2018}. We checked, however, that in the hydrodynamic approximation
the latter correspond to the hybridized surface-bulk modes with large frequencies. In addition, the open contours at
large enough negative $k_x$ (or $q_y$ in the notation of Ref.~\cite{Andolina:2018}) in Fig.~2(a) of
Ref.~\cite{Andolina:2018} could, presumably, stem from the hybridization of the gapped and gapless
modes.

%%%%%%%%%%%%%%%%%%
%%%%%%%%%%%%%%%%%%

In general, we identify two qualitative features that could be used to analyze the effects of the Fermi arcs on
the surface collective modes in Weyl semimetals. First of all, unlike the conventional surface plasmons with
the frequency given in Eq.~(\ref{SP-no-FA-omega}), the Fermi arc surface plasmons are described by a strongly
anisotropic dispersion relation. Also, unlike the gapped plasmon mode, which exists even in the absence of the
surface Fermi arc states, the new gapless collective mode appears only when the topological surface states
are taken into account. Therefore, if experimentally observed, the latter mode can be used to extract the information about
the separation between the Weyl nodes as well as the dispersion relation of the Fermi arcs.

Experimentally, the anisotropy induced by the surface states can be probed using the near-field optical
spectroscopy (for a review, see Ref.~\cite{Basov-rev:2016}), as well as the momentum-resolved electron
energy loss spectroscopy (see, e.g., Ref.~\cite{Wang:1995}). Because of a possible interference between
the surface modes from different pairs of Weyl nodes \cite{Song:2017}, the most suitable materials are the
Weyl semimetals with a single pair of nodes.
Therefore, the magnetic Heusler compounds with a broken TR symmetry \cite{Hasan-magnetic:2016,Cava-Bernevig:2016}
might be promising candidates for the study of the surface collective modes.

\section{Summary}
\label{sec:Summary}

In this paper, we proposed that the Weyl semimetals with a broken TR symmetry may possess a hydrodynamic
regime with a nontrivial interplay between the bulk electron fluid and the fluid formed by the surface Fermi arc quasiparticles.
The hydrodynamic equations for the latter are derived from the kinetic theory under the assumptions that the electron-electron scattering rate
dominates over the electron-impurity and electron-phonon ones.
Further, we considered only the case where the hydrodynamic regime is achieved for both surface and
bulk quasiparticles of the semimetal. In principle, however, the regime where the electron fluid is formed only on the surface but not in the bulk could be also realized.
Such a scenario is likely to lead to unique features and deserves a separate study.

In the proposed two-fluid framework, we studied the role of the Fermi arc fluid on the bulk flow and on the spectrum
of surface collective modes. For simplicity, we assumed that the surface fluid is inviscid and couples to the bulk
via the phenomenological inflow and outflow terms. The latter describe the Fermi arc dissipation into the
bulk and the transitions from the bulk to the surface states. We found that the Fermi arcs modify the
boundary conditions for the bulk electron fluid. Depending on the rate of the surface-bulk transitions
as well as the value of the chiral shift, the bulk fluid velocity could change significantly near the
boundaries. When the electrons are transferred to the surface at a greater rate than to the bulk, the bulk
fluid could be dragged by the surface one. Such a regime, however, is characterized by large surface
flow velocities at which the hydrodynamic description may become ill defined. On the other hand, an
unconventional increase of the bulk fluid flow near the boundaries is seen when the surface to bulk
transitions dominate.
Such a manifestation of the Fermi arc flow could be, in principle, observed via the decrease of the resistivity in samples of small width.

In this study, we also demonstrated that the Fermi arcs profoundly affect the surface collective modes in the hydrodynamic regime. In
particular, we found that the dispersion relations of the surface plasmons become anisotropic in momentum space.
This is in contrast to the conventional surface plasmons with the isotropic dispersion.
The origin of the anisotropy is the dispersion relation of the surface Fermi arc quasiparticles.
In general, we identified two types of surface modes. While one of them is a gapped surface plasmon hybridized with the Fermi arc oscillations, the other is a gapless mode triggered
exclusively by the surface states. Similarly to the usual surface acoustic plasmons~\cite{Pitarke:2004}, the gapless Fermi arc
mode has a linear dispersion relation, but it is sensitive to the sign of the wave-vector component along
the direction of the Fermi arc velocity. While our results agree qualitatively with those in
Refs.~\cite{Song:2017,Andolina:2018}, we argue that the only true surface plasmon modes are those with the
closed elliptic contours of constant frequency.

In passing, let us discuss a few limitations of this study. The hydrodynamic model proposed in this
paper is phenomenological and the underlying reasons for the fluid formation have not been rigorously addressed.
In addition, the reliable estimate of the hydrodynamic window, i.e., the parameter region where the electron fluid can be formed, is still lacking for many experimentally realized Weyl semimetals.
In our analysis, we used a simplified model for the Fermi arcs without any curvature. While we believe
that the results will remain qualitatively the same for slightly curved arcs, the precise role of a nonzero
curvature should be addressed in the future. In our study of the surface collective modes, we also neglected
the viscosity and dissipation effects, which could
be rigorously taken into account via nonlocal corrections as in Ref.~\cite{Andolina:2018}.
In the future, it would be interesting to address also the effects of multiple pairs of Weyl nodes and, therefore, several Fermi arcs on the formation of the surface fluid. Such an investigation is beyond the scope of this study, however.

\begin{acknowledgments}
The work of E.V.G. was partially supported by the Program of Fundamental Research of the
Physics and Astronomy Division of the National Academy of Sciences of Ukraine.
The work of V.A.M. and P.O.S. was supported by the Natural Sciences and Engineering Research Council of Canada.
The work of I.A.S. was supported by the U.S. National Science Foundation under Grant PHY-1713950.
P.O.S. is grateful for the hospitality of Nordita during the program ``Quantum Anomalies and Chiral Magnetic
Phenomena", where a part of the study was done, as well as appreciates the discussion with
Prof.~V.~Juri\v{c}i\'{c} at the initial stage of the project.
\end{acknowledgments}

\appendix

\section{Derivation of the Fermi arc hydrodynamics}
\label{sec:App-derivation}

In this appendix, we present the technical details of derivation of the hydrodynamic equation for the
Fermi arc surface states. We utilize a simple model of a time-reversal symmetry breaking Weyl semimetal
with two Weyl nodes separated in momentum space by $2b$ along the $z$ direction, where $b$ is the
magnitude of the chiral shift. The semimetal is finite along the $y$ direction and infinite in the other two.

\subsection{Kinetic theory}
\label{sec:App-derivation-kinetic}

We follow the standard approach \cite{Landau:t10,Huang-book} of deriving the
hydrodynamic equation from the kinetic theory. In the presence of an electric field $\mathbf{E}$, the
kinetic equation reads as
\begin{equation}
\partial_t f^{\rm (FA)} -e\mathbf{E}\cdot \partial_{\mathbf{p}} f^{\rm (FA)} +\mathbf{v}_p^{\rm (FA)} \cdot\bm{\nabla} f^{\rm (FA)} =I^{\rm (FA)}_{\rm coll},
\label{derivation-FA-KT-eq}
\end{equation}
where $-e$ is the electron charge, $\mathbf{p}=\left(p_x, p_z\right)$ is the momentum of the surface quasiparticles
and $I^{\rm (FA)}_{\rm coll}$ denotes the collision integral, whose explicit form will be discussed later.

Since hydrodynamics assumes a local equilibrium, we take the distribution function in the following form:
\begin{equation}
f^{\rm (FA)} = \delta(y-y_s) \frac{1}{1+\rm{exp}\left(\frac{\epsilon_{p}^{\rm (FA)}-(\mathbf{u}^{\rm (FA)}\cdot\mathbf{p}) -\mu}{T}\right)},
\label{derivation-FA-Fermi-Dirac}
\end{equation}
where $y_s$ denotes the surface coordinate, $s=\pm$ denotes the bottom ($+$) or top ($-$) surface, $\mathbf{u}^{\rm (FA)}$ is the local fluid velocity of the surface
Fermi arc states, $\mu$ is the electric chemical potential, and $T$ is temperature. For a slab of finite thickness,
$y_{-}=L_y$ and $y_{+}=0$ denote the top and bottom surfaces, respectively. Here, we assume that the Fermi
arcs are strongly localized at the surface and the dependence of the distribution function on the transverse
coordinate can be modeled by the $\delta$ function.

The quasiparticle energy for the surface states reads as
\begin{equation}
\label{derivation-FA-epsilon_p}
\epsilon_{p}^{\rm (FA)}= sv_Fp_x,
\end{equation}
where $v_F$ is the Fermi velocity. (For the
derivation of the Fermi arcs and their dispersion relation see, e.g.,
Refs.~\cite{Murakami:2014,Gorbar:2016aov,Gorbar:2017lnp}.) The corresponding quasiparticle velocity
is given by
\begin{equation}
\mathbf{v}_p^{\rm (FA)}= \partial_\mathbf{p}\epsilon_{p}^{\rm (FA)}
=sv_F\hat{\mathbf{x}},
\label{derivation-FA-v_p}
\end{equation}
where $\hat{\mathbf{x}}$ is the unit vector in the $x$ direction. Since the Fermi arc quasiparticles move
only along the $x$ axis, it is reasonable to assume that the surface hydrodynamic motion is also possible
only along that axis, i.e., $\mathbf{u}^{\rm (FA)}\parallel \hat{\mathbf{x}}$. As we will show in Appendix~\ref{sec:App-FA-charge-current}, this is further
justified by the fact that the Fermi arc electric current can only flow along the $x$ direction.

In the case of small fluid velocities, we can use the following expansion for the distribution function:
\begin{equation}
f^{\rm (FA)} \approx f^{\rm (FA,0)} -p_x u_x^{\rm (FA)} \frac{\partial f^{\rm (FA,0)}}{\partial \epsilon_p^{\rm (FA)}},
\label{derivation-FA-f-exp}
\end{equation}
where
\begin{equation}
f^{\rm (FA,0)} = \delta(y-y_s)\frac{1}{1+e^{(\epsilon_p^{\rm (FA)}-\mu)/T}}
\label{derivation-FA-f0}
\end{equation}
is the distribution function of the Fermi arc quasiparticles in global equilibrium.

\subsection{The Euler equation for the Fermi arc fluid}
\label{sec:App-derivation-Euler}

In order to derive the Euler equation, we multiply Eq.~(\ref{derivation-FA-KT-eq}) by the $x$ component
of the momentum $p_x$ and integrate over $p_x$. [It should be noted that, because of the dispersion relation (\ref{derivation-FA-epsilon_p}), there is no independent energy conservation equation.]

The integration of the first term in Eq.~(\ref{derivation-FA-KT-eq}) leads to the following result:
\begin{eqnarray}
\label{derivation-FA-Euler-time-start}
\int \frac{d^2p}{(2\pi \hbar)^2} p_x \partial_t f^{\rm (FA)} &=&
\int \frac{d^2p}{(2\pi \hbar)^2} p_x \partial_t \left( f^{\rm (FA,0)} -p_x u_x^{\rm (FA)} \frac{\partial f^{\rm (FA,0)}}{\partial \epsilon_p^{\rm (FA)} }\right)
= -\partial_t \sum_{\rm p, a} \frac{s F T^2}{2\pi v_F^2 \hbar^2} \mathrm{Li}_2\left(-e^{\mu/T}\right)
\nonumber\\
&-& \sum_{\rm p, a} \partial_t \frac{F T^2 u_x^{\rm (FA)}}{\pi v_F^3 \hbar^2} \left[\frac{\pi^2}{6} -\mathrm{Li}_2\left(1+e^{\mu/T}\right) -\left(\frac{\mu}{T} +i \pi \right) \ln{\left(1+e^{\mu/T}\right)} \right]
 \nonumber\\
&=& \partial_t \frac{sF}{4\pi v_F^2 \hbar^2} \left(\mu^2 +\frac{\pi^2T^2}{3}\right) \left(1+\frac{2u_x^{\rm (FA)}}{s v_F}\right),
\end{eqnarray}
where we took into account the small fluid velocity expansion in Eq.~(\ref{derivation-FA-f-exp}) and
use the formulas in Appendix~\ref{sec:App-ref}. {Further, $\sum_{\rm p, a}$ denotes
the summation over particles (electrons) and antiparticles (holes). It should be noted that $\mu\to-\mu$
and $e\to -e$ for holes and the limits of integration over $p_x$ depends on the boundary label $s$, i.e.,
$\int_0^{s\infty} dp_x$. The overall coefficient $F$ is defined by the integration over the length of the Fermi arc,
i.e.,
\begin{equation}
\label{derivation-FA-mod-Euler-F-def}
F = \int_{-\hbar b}^{\hbar b} \frac{dp_z}{2\pi}   = \frac{\hbar b}{\pi}.
\end{equation}
The integral with the term containing the electric field in Eq.~(\ref{derivation-FA-KT-eq}) can be calculated in a similar way. The result reads as
\begin{eqnarray}
\label{derivation-FA-Euler-momentum-start}
-e\int \frac{d^2p}{(2\pi \hbar)^2} p_x \left(\mathbf{E}\cdot \partial_{\mathbf{p}}\right) f^{\rm (FA)}
&=&\sum_{\rm p, a} \frac{e}{2\pi \hbar^2} \frac{E_x F T}{v_F}\ln{\left(1+e^{\mu/T}\right)}
+ \sum_{\rm p, a} \frac{se}{2\pi \hbar^2} \frac{u_x^{\rm (FA)} E_x F T}{v_F^2}\ln{\left(1+e^{\mu/T}\right)}
\nonumber\\
&=& \frac{e E_xF\mu}{2\pi v_F \hbar^2} \left(1+ \frac{u_x^{\rm (FA)}}{sv_F}\right).
\end{eqnarray}
The term with the spatial derivatives gives rise to the following result:
\begin{eqnarray}
\label{derivation-FA-Euler-gradient-start}
\int \frac{d^2p}{(2\pi \hbar)^2} p_x \left(\mathbf{v}_p^{\rm (FA)} \cdot\bm{\nabla}\right) f^{\rm (FA)}
&=& - \sum_{\rm p, a}  sv_F\partial_x \frac{F T^2}{2\pi v_F^2\hbar^2} \mathrm{Li}_2\left(-e^{\mu/T}\right)
\nonumber\\
&-& \sum_{\rm p, a} sv_F\partial_x \frac{F T^2 u_x^{\rm (FA)}}{\pi v_F^3 \hbar^2} \left[\frac{\pi^2}{6}
-\mathrm{Li}_2\left(1+e^{\mu/T}\right) -\left(\frac{\mu}{T} +i \pi\right) \ln{\left(1+e^{\mu/T}\right)} \right]
\nonumber\\
&=& \partial_x \frac{F}{4\pi v_F \hbar^2} \left(\mu^2 +\frac{\pi^2 T^2}{3}\right) \left(1+2\frac{u_x^{\rm (FA)}}{sv_F}\right).
\end{eqnarray}
By collecting all contributions together, we finally arrive at the following Euler equation for the Fermi arc fluid:
\begin{equation}
\label{derivation-FA-Euler-final}
\left(\partial_t+sv_F \partial_x\right) \frac{s F}{4\pi v_F^2\hbar^2} \left(\mu^2 +\frac{\pi^2 T^2}{3}\right)
\left(1+2\frac{u_x^{\rm (FA)}}{sv_F}\right) +\frac{e\mu F}{2\pi v_F \hbar^2} \left(1+\frac{u_x^{\rm (FA)}}{s v_F}\right) E_x
= I^{\rm (FA)}.
\end{equation}

\subsection{Transfer term}
\label{sec:App-transfer-term}

Here, we present the derivation of the transfer term on the right-hand side of the Euler equation
(\ref{derivation-FA-Euler-final}). In general, it may contain two different parts: one describing the surface
to bulk transitions and the other describing the inflow from the bulk. By recalling that the dissipation
of the Fermi arcs is primarily due to the surface to bulk scatterings~\cite{Gorbar:2016aov}, the first
part of $I^{\rm (FA)}$ can be obtained by using the relaxation time approximation as follows:
\begin{eqnarray}
\label{FA-bulk-connection-FA-to-Bulk}
&&-\int \frac{d^2p}{(2\pi \hbar)^2} p_x \frac{f^{\rm (FA)}-f^{\rm (FA,0)}}{\tau_{sb}} =
\int \frac{d^2p}{(2\pi \hbar)^2} p_x  \frac{p_xu_x^{\rm (FA)}}{\tau_{sb}} \frac{\partial f^{\rm (FA,0)}}{\partial \epsilon_p^{\rm (FA)}} \nonumber\\
&&=  \frac{1}{\tau_{sb}} \sum_{\rm p, a}  \frac{F T^2 u_x^{\rm (FA)}}{\pi v_F^3 \hbar^2} \Big[\frac{\pi^2}{6}  -\mathrm{Li}_2\left(1+e^{\mu/T}\right) -\left(\frac{\mu}{T} +i \pi\right) \ln{\left(1+e^{\mu/T}\right)} \Big] \nonumber\\
&&= -\frac{u_x^{\rm (FA)}}{\tau_{sb}} \frac{F}{2\pi v_F^3 \hbar^2} \left(\mu^2 +\frac{\pi^2T^2}{3}\right).
\end{eqnarray}
The term describing the bulk to surface transitions, on the other hand, can be calculated by using the method in
the Supplemental Material of Ref.~\cite{Gorbar:2017vph}. Its explicit form reads as
\begin{equation}
\label{FA-bulk-connection-Bulk-to-FA}
\lambda_B\frac{w u_x(y_s)}{\tau_{bs}v_F^2},
\end{equation}
where $u_x(y_s)$ is the bulk fluid velocity on the surface, $w$ in the bulk enthalpy density, $\lambda_B$
is the dimensional coefficient, and $\tau_{bs}$ is the relaxation time describing bulk to surface transitions.
It might be more convenient to parametrize the bulk inflow in terms of a single overall coefficient
$\alpha=\lambda_B /(v_F\tau_{bs})$. Then, the final expression for the transfer term $I^{\rm (FA)}$
takes the form as in Eq.~(\ref{model-Icoll}) in the main text.

\subsection{Electric charge and current densities of Fermi arcs}
\label{sec:App-FA-charge-current}

For completeness, we present the explicit expressions for the electric charge and current densities for Fermi arc quasiparticles.
The corresponding expressions can be obtained by using the kinetic theory, i.e.,
\begin{eqnarray}
\label{derivation-FA-currents-rho-FA}
\rho^{\rm (FA)}&=& -\sum_{\rm p, a} e \int \frac{d^2p}{(2\pi \hbar)^2} f^{\rm (FA)} = -\sum_{\rm p, a} e \int \frac{d^2p}{(2\pi \hbar)^2} \left[f^{\rm (FA,0)} - p_xu_x^{\rm (FA)} \frac{\partial f^{\rm (FA,0)}}{\partial \epsilon_p^{\rm (FA)}}\right] \nonumber\\
&=&- \sum_{\rm p, a} \frac{seF}{2\pi \hbar^2} \left[\frac{T}{sv_F} \ln{\left(1+e^{\mu/T}\right)} +u_{x}^{\rm (FA)} \frac{T}{s^2v_F^2} \ln{\left(1+e^{\mu/T}\right)} \right] = -\frac{e \mu F}{2\pi v_F \hbar^2} \left(1+\frac{u_x^{\rm (FA)}}{sv_F}\right)
\end{eqnarray}
and
\begin{eqnarray}
\label{derivation-FA-currents-J-FA}
\mathbf{J}^{\rm (FA)}&=& -\sum_{\rm p, a} e \int \frac{d^2p}{(2\pi \hbar)^2} \mathbf{v}_p^{\rm (FA)} f^{\rm (FA)} = -\sum_{\rm p, a} e s v_F \hat{\mathbf{x}} \int \frac{d^2p}{(2\pi \hbar)^2} \left[f^{\rm (FA,0)} - p_xu_x^{\rm (FA)} \frac{\partial f^{\rm (FA,0)}}{\partial \epsilon_p^{\rm (FA)}}\right] \nonumber\\
&=&- \sum_{\rm p, a} v_F \hat{\mathbf{x}}\frac{eF}{2\pi \hbar^2} \left[\frac{T}{sv_F} \ln{\left(1+e^{\mu/T}\right)} +u_{x}^{\rm (FA)} \frac{T}{s^2v_F^2} \ln{\left(1+e^{\mu/T}\right)} \right] = - s\hat{\mathbf{x}} \frac{e\mu F}{2\pi \hbar^2} \left(1+\frac{u_x^{\rm (FA)}}{sv_F}\right),
\end{eqnarray}
respectively.

\section{Polylogarithm functions}
\label{sec:App-ref}

In this appendix, we present several definitions and identities for the polylogarithm functions
used in the derivation of the Euler equation. By making use of the short-hand notation $f^{(0)} =1/[e^{(v_Fp-\mu)/T}+1]$, it is straightforward to derive the following formulas:
\begin{eqnarray}
\int_0^{\infty} dp\, p^{n}  f^{(0)}
&=& -\frac{T^{n+1} \Gamma(n+1) }{v_F^{n+1}}  \mbox{Li}_{n+1}\left(-e^{\mu/T}\right),
\qquad n\geq 0,
\label{integral-3a} \\
\int_0^{\infty} dp\, p^{n} \frac{\partial f^{(0)}}{\partial p}
&=& \frac{T^{n} \Gamma(n+1) }{v_F^{n}}  \mbox{Li}_{n}\left(-e^{\mu/T}\right),
\qquad n\geq 0,
\label{integral-3b}
\end{eqnarray}
where $\mbox{Li}_{n}(x)$ is the polylogarithm function. The polylogarithm functions of order $n=0$ and $1$ can be expressed in terms of elementary functions, i.e.,
\begin{eqnarray}
\mbox{Li}_{0}\left(-e^{x}\right) &=& -\frac{1}{1+e^{-x}}, \\
\mbox{Li}_{1}\left(-e^{x}\right) &=& -\ln{\left(1+e^{x}\right)}.
\label{App-polylog}
\end{eqnarray}
Also, the following identities are useful
\begin{eqnarray}
\label{App-polylog-sum-be}
&&\mbox{Li}_{0} (-e^{x}) +\mbox{Li}_{0} (-e^{-x})   = -1,\\
&&\mbox{Li}_{1} (-e^{x}) -\mbox{Li}_{1} (-e^{-x})   = -x,\\
&&\mbox{Li}_{2} (-e^{x}) +\mbox{Li}_{2} (-e^{-x})   = -\frac{1}{2}\left(x^2+\frac{\pi^2}{3}\right),\\
&&\mbox{Li}_{2} (1+e^{x}) + \mbox{Li}_{2} (1+e^{-x}) +i\pi \left[\ln{\left(1+e^{x}\right)} + \ln{\left(1+e^{-x}\right)}\right] = \frac{1}{2} \left[\pi^2 -\ln^2{\left(e^{x}\right)}\right].
\label{App-polylog-sum-ee}
\end{eqnarray}

\end{document}